\documentclass[12pt,a4paper]{article}
\usepackage{epsfig}
\begin{document}
\title{Production of the top-pions from the higgsless--\\top-Higgs model at the LHC}
\author{Chong-Xing Yue and  Yi-Qun Di  \\
{\small Department of Physics, Liaoning  Normal University, Dalian,
116029 P. R. China}
\thanks{E-mail:cxyue@lnnu.edu.cn}}
\date{\today}
\maketitle

\begin{abstract}The top-pions($\pi_{t}^{0,\pm}$) predicted by extra
dimensional descriptions of the topcolor scenario have similar
feature with those in four dimensional topcolor scenario, which have
large Yukawa couplings to the third generation quarks. In the
context of the higgsless--top-Higgs(HTH) model, we discuss the
production of these new particles at the CERN Large Hadron
Collider(LHC) via various suitable mechanisms (gluon-gluon fusion,
bottom-bottom fusion, gluon-bottom fusion, and the usual Drell-Yan
processes) and estimate their production rates. We find that, as
long as the top-pions are not too heavy, they can be abundantly
produced at the LHC. The possible signatures of these new particles
might be detected at the LHC experiments.

\hspace{5mm}

\vspace{1cm}


\end {abstract}

\newpage
\noindent{\bf I. Introducton}

To completely avoid the problems arising from the elementary Higgs
field in the standard model(SM), various models for dynamical
electroweak symmetry breaking(EWSB) have been proposed, among which
the topcolor scenario is attractive because it can explain the large
top mass and provide a possible EWSB mechanism[1]. The common
feature of this kind of models is that topcolor interactions make
small contributions to EWSB and give rise to most part of the top
mass, while EWSB is mainly induced by technicolor or other strong
interactions.

Recently, the model building along with the TeV-scale extra
dimensional scenario[2] has been widely studied. Considering the
gauge symmetry breaking can be easily achieved by imposing
appropriate boundary conditions[3], the higgsless theory[4] was
proposed and the topcolor scenario was reconsidered in extra
dimensions[5,6]. For extra dimensional descriptions of the topcolor
scenario[5], there are two separate 5 dimensional anti de Sitter
spaces($AdS'_{5}s$) with each sector having its own corresponding
infrared(IR) brane. The light fermions propagate in an $AdS_{5}$
sector, while the third generation quarks would propagate in other
$AdS_{5}$ sector, and sources for EWSB are localized on both of
these IR branes. Using the appropriate boundary and matching
conditions, there are three possibilities for EWSB:
Higgs--top-Higgs, higgsless--top-Higgs, and higgsless--higgsless.
For the higgsless--top-Higgs case, most of EWSB comes from the
higgsless sector, and the top quark gets its mass from a top-Higgs
on the other IR brane.

Similarly to the topcolor scenario in four dimensional space, the
higgsless--top-Higgs (HTH) model[5] predicts the existence of
isotriplet scalars, called top-pions($\pi_{t}^{0,\pm}$), which have
large Yukawa couplings to the third generation quarks and sensibly
small couplings to the massive gauge bosons and light quarks. The
top-pions get masses at one-loop from gauge interactions, which can
propagate from one boundary to the other. The mass scale is set by
the cutoff scale on the IR brane, so the top-pions should be quite
heavy and can not be abundantly produced at the present high energy
experiments. The aim of this paper is to investigate production of
these new particles ($\pi_{t}^{0,\pm}$) at the CERN Large Hadron
Collider(LHC) with $\sqrt{s}=14TeV$ and see whether the possible
signatures of the HTH model can be detected in the near future LHC
experiments.

In the rest of this paper, we will give our results in detail. The
couplings of the top-pions $\pi_{t}^{0,\pm}$ to the ordinary
particles are given in Sec.II, and the decay widths of the possible
decay modes are also estimated in this section. Sections III and IV
are devoted to the computation of the production cross sections of
$\pi_{t}^{0}$ and $\pi_{t}^{\pm}$ at the LHC, respectively. Some
phenomenological analysis are also included in these sections. Our
conclusions are given in Sec.V.

\noindent{\bf II. The possible decay modes of the top-pions
$\pi_{t}^{0,\pm}$}

For the HTH model[5], there are two $AdS_{5}$ spaces intersecting
along a codimension one surface(Planck brane) that would serve as a
ultraviolet(UV) cutoff of the two $AdS_{5}$ spaces. The two
$AdS_{5}$ spaces, denoted as $AdS_{5}^{w}$ and $AdS_{5}^{t}$, are
characterized by their own curvature scale $R_{w}$ and $R_{t}$,
respectively. The common UV boundary for each brane is located at
the point $Z=R_{w}$ or $R_{t}$ in the coordinate system and each
$AdS_{5}$ space is also cut by an IR boundary located at $Z=R'_{w}$
or $R'_{t}$. The IR brane at $R'_{w}$ has a higgsless boundary
condition and is responsible for most of EWSB, while the top-Higgs
$h_{t}^{0}$ on the IR brane at $R'_{t}$ makes very small
contributions to EWSB. Thus, for the HTH model, there is the
following relation:
\begin{equation}
\nu^{2}=\nu^{2}_{w}+\nu^{2}_{t},
\end{equation}
where $\nu=246GeV$ is the electroweak scale, $\nu_{w}$ represents
the contributions of the higgsless sector to EWSB. The vacuum
expectation value(VEV) $\nu_{t}$ of the top-Higgs produces all of
the top mass. To produce the large top mass $m_{t}$, the VEV
$\nu_{t}$ should be very small. Thus, the top-pions have large
Yukawa couplings to the third generation quarks.

The coupling constants of the top-pions $\pi_{t}^{0,\pm}$ to the
third generation quarks can be written as[5]:
\begin{equation}
\pi_{t}^{0}t\overline{t}: \frac{m_{t}}{\nu_{t}},\ \ \
\pi_{t}^{0}b\overline{b}: \frac{m_{b}}{\nu_{t}},
\end{equation}
\begin{equation}
\pi_{t}^{-}t\overline{b}:
\frac{e}{\sqrt{2}S_{W}m_{W}}\frac{\nu}{\nu_{t}}[m_{t}P_{R}+m_{b}P_{L}],
\end{equation}
where $S_{W}=sin\theta_{W}, \theta_{W}$ is the Weinberg angle, and
$P_{L(R)}=(1\mp\gamma_{5})/2$ is the left(right)-handed projection
operator. Certainly, there might be the flavor changing(FC)
couplings $\pi_{t}^{0}\overline{t}c$ and $\pi_{t}^{-}\overline{b}c$.
However, for the HTH model, the flavor physics is generated on the
Planck brane via mixing in non-diagonal localized kinetic terms, one
can choose the free parameters to avoid the presence of the FC
couplings $\pi_{t}^{0}\overline{t}c$ and $\pi_{t}^{-}\overline{b}c$.
Furthermore, considering the electorweak precision measurement
constraints, the top-pions should be quite heavy. Through out this
paper, we will take the top-pion mass in the range of
$400GeV\sim800GeV$. In this case, the main decay channels are
$\pi_{t}^{0}\rightarrow t\overline{t}$ and $\pi_{t}^{\pm}\rightarrow
tb$. Thus, we will not consider the FC couplings of
$\pi_{t}^{0,\pm}$ in this paper. Compared to the SM Higgs boson, the
couplings of $\pi_{t}^{0}$ to the third generation quarks are
enhanced by the factor $\nu/\nu_{t}$, while the tree-level couplings
of $\pi_{t}^{0}$ to the electroweak gauge bosons $W$ and $Z$ are
suppressed by the factor $\nu_{t}/\nu$:
\begin{equation}
\pi_{t}^{0}ZZ:\frac{\nu_{t}}{\nu}\frac{em_{Z}}{S_{W}C_{W}}g_{\mu\nu},
\ \ \
\pi_{t}^{0}W^{+}W^{-}:\frac{\nu_{t}}{\nu}\frac{em_{W}}{S_{W}}g_{\mu\nu}.
\end{equation}

For the neutral top-pion $\pi_{t}^{0}$ decays, we will focus our
attention on the decay modes: $t\overline{t}, b\overline{b},
W^{+}W^{-}, ZZ, gg$ and $\gamma\gamma$. The expressions of the decay
widths for these decay channels can be written as:
\begin{equation}
\Gamma(\pi_{t}^{0}\rightarrow
t\overline{t})=\frac{3m^{2}_{t}}{8\pi\nu_{t}^{2}}m_{\pi_{t}}
(1-\frac{4m^{2}_{t}}{m^{2}_{\pi_{t}}})^{\frac{3}{2}},
\end{equation}
\begin{equation}
\Gamma(\pi_{t}^{0}\rightarrow
b\overline{b})=\frac{3m^{2}_{b}}{8\pi\nu_{t}^{2}}m_{\pi_{t}}
(1-\frac{4m^{2}_{b}}{m^{2}_{\pi_{t}}})^{\frac{3}{2}},
\end{equation}
\begin{equation}
\Gamma(\pi_{t}^{0}\rightarrow
W^{+}W^{-})=\frac{\alpha_{e}m^{3}_{\pi_{t}}}{16S_{W}^{2}m_{W}^{2}}
(\frac{\nu_{t}}{\nu})^{2}
(1-\frac{4m^{2}_{W}}{m^{2}_{\pi_{t}}})^{\frac{1}{2}}(1-4\frac{m^{2}_{W}}{m^{2}_{\pi_{t}}}
+12\frac{m^{4}_{W}}{m^{4}_{\pi_{t}}}),
\end{equation}
\begin{equation}
\Gamma(\pi_{t}^{0}\rightarrow
ZZ)=\frac{\alpha_{e}m^{3}_{\pi_{t}}}{32S_{W}^{2}m_{W}^{2}}
(\frac{\nu_{t}}{\nu})^{2}
(1-\frac{4m^{2}_{Z}}{m^{2}_{\pi_{t}}})^{\frac{1}{2}}(1-4\frac{m^{2}_{Z}}{m^{2}_{\pi_{t}}}
+12\frac{m^{4}_{Z}}{m^{4}_{\pi_{t}}}),
\end{equation}
\begin{equation}
\Gamma(\pi_{t}^{0}\rightarrow gg)
=\frac{2\alpha^{2}_{s}m^{3}_{\pi_{t}}m^{4}_{t}}{9\nu^{2}_{t}\pi^{3}}
[(1-4\frac{m^{2}_{t}}{m^{2}_{\pi_{t}}})C_{0}(m_{t}^{2})-\frac{2}{m^{2}_{\pi_{t}}}]^{2},
\end{equation}
\begin{eqnarray}
\Gamma(\pi_{t}^{0}\rightarrow \gamma\gamma)
&=&\frac{\alpha^{2}_{e}m^{3}_{\pi_{t}}}{16\pi^{3}}
\{\frac{4}{3\nu^{2}_{t}}[m^{2}_{t}(1-4\frac{m^{2}_{t}}{m^{2}_{\pi_{t}}})C_{0}(m_{t}^{2})
-\frac{2m^{2}_{t}}{m^{2}_{\pi_{t}}}]\nonumber\\
&&+\frac{\nu^{2}_{t}}{\nu^{4}}[\frac{1}{2}+3\frac{m^{2}_{W}}{m^{2}_{\pi_{t}}}-3m^{2}_{W}
(1-2\frac{m^{2}_{W}}{m^{2}_{\pi_{t}}})C_{0}(m^{2}_{W})]\}^{2}.
\end{eqnarray}
The expression of the three-point scalar integral $C_{0}(m^{2}_{i})$
can be written as
$C_{0}(m^{2}_{i})=C_{0}(m^{2}_{\pi_{t}},0,0,m^{2}_{i},m^{2}_{i},m^{2}_{i})$[7].
For the decay channels $\pi^{0}_{t}\rightarrow gg$ and
$\gamma\gamma$, we have neglected the contributions of the light
quarks.

To obtain numerical results, we need to specify the relevant SM
input parameters. We take the SM parameters as $m_{t}=172.7GeV$[8],
$S^{2}_{W}=0.2315, \alpha_{e}=1/128.8, \alpha_{s}=0.112,
m_{Z}=91.187GeV$, and $m_{W}=80.425GeV$[9]. Except for these SM
input parameters, the partial decay widths are dependent on the two
free parameters $\nu_{t}$ and $m_{\pi_{t}}$. To make that the HTH
model generates enough large top mass and satisfies the electroweak
precision constraints, the top-Higgs VEV $\nu_{t}$ should be very
small and the top-pion should be quite heavy[5]. As numerical
estimation, we will take $\nu_{t}\leq50GeV(\nu_{t}/\nu\leq1/5)$ and
assume that the top-pion mass $m_{\pi_{t}}$ is in the range of
$400GeV\sim800GeV$.

The branching ratios of the prominent decay channels for the neutral
top-pion $\pi^{0}_{t}$ are pictured in Fig.1 as functions of the
top-pion mass $m_{\pi_{t}}$ for $\nu_{t}/\nu\approx1/5$, in which we
have multiplied the factors 100 and 10 to the branching ratios
$Br(\pi^{0}_{t}\rightarrow b\overline{b})[Br(\pi^{0}_{t}\rightarrow
W^{+}W^{-}), Br(\pi^{0}_{t}\rightarrow ZZ)]$ and
$Br(\pi^{0}_{t}\rightarrow gg)$, respectively. Since the value of
the branching ratio $Br(\pi^{0}_{t}\rightarrow \gamma\gamma)$ is
largely smaller than that of $Br(\pi^{0}_{t}\rightarrow
t\overline{t})$, so we have not shown $Br(\pi^{0}_{t}\rightarrow
\gamma\gamma)$ in Fig.1. From Fig.1 we can see that, although the
decay widths for all of the possible decay channels increase as
$m_{\pi_{t}}$ increasing, the main decay modes of $\pi^{0}_{t}$ are
$t\overline{t}$ and $gg$. For $400GeV\leq m_{\pi_{t}}\leq 800GeV$,
the value of $Br(\pi^{0}_{t}\rightarrow gg)$ is in the range of
$1.33\%\sim0.34\%$. Certainly, the values of the branching ratios
for various possible modes vary as the parameter $\nu_{t}/\nu$
varying. For example, the value of the branching ratio
$Br(\pi^{0}_{t}\rightarrow gg)$ increases from $1.33\%$ to $1.35\%$
as $\nu_{t}/\nu$ decreasing from $1/5$ to $1/8$. Thus the neutral
top-pion $\pi^{0}_{t}$ might be abundantly produced at the LHC via
the subprocess $gg\rightarrow\pi^{0}_{t}$. The values of the
branching ratios for all possible decay modes are given in Table I
for three values of the top-pion mass $m_{\pi_{t}}$ and the
top-Higgs $VEV$ $ \nu_{t}$.

\begin{center}
{
\begin{small}
\begin{tabular}{|c|c||c|c|c|c|c|c|}
\hline
$\nu_{t}(GeV)$&$m_{\pi_{t}}(GeV)$&$Br(t\overline{t})$&$Br(b\overline{b})
$&$Br(W^{+}W^{-})$&$Br(ZZ)$&$Br(gg)$\\
\hline
$$&$400$&$96.69\%$&$0.56\%$&$0.97\%$&$0.45\%$&$1.33\%$\\
\cline{2-2}\cline{3-3}\cline{4-4}\cline{5-5}\cline{6-6}\cline{7-7}
$50$&$600$&$98.63\%$&$0.12\%$&$0.54\%$&$0.26\%$&$0.46\%$\\
\cline{2-2}\cline{3-3}\cline{4-4}\cline{5-5}\cline{6-6}\cline{7-7}
$$&$800$&$98.47\%$&$0.09\%$&$0.74\%$&$0.36\%$&$0.34\%$\\
\hline
$$&$400$&$97.50\%$&$0.57\%$&$0.40\%$&$0.19\%$&$1.34\%$\\
\cline{2-2}\cline{3-3}\cline{4-4}\cline{5-5}\cline{6-6}\cline{7-7}
$40$&$600$&$90.09\%$&$0.12\%$&$0.22\%$&$0.11\%$&$0.46\%$\\
\cline{2-2}\cline{3-3}\cline{4-4}\cline{5-5}\cline{6-6}\cline{7-7}
$$&$800$&$99.11\%$&$0.09\%$&$0.30\%$&$0.15\%$&$0.34\%$\\
\hline
$$&$400$&$97.89\%$&$0.57\%$&$0.13\%$&$0.06\%$&$1.35\%$\\
\cline{2-2}\cline{3-3}\cline{4-4}\cline{5-5}\cline{6-6}\cline{7-7}
$30$&$600$&$99.31\%$&$0.12\%$&$0.07\%$&$0.03\%$&$0.46\%$\\
\cline{2-2}\cline{3-3}\cline{4-4}\cline{5-5}\cline{6-6}\cline{7-7}
$$&$800$&$99.42\%$&$0.09\%$&$0.10\%$&$0.05\%$&$0.34\%$\\
\hline\end{tabular}
\end{small} }\end{center}
\hspace{0.3cm} Table I: The values of leading branching ratios for
three values of the top-Higgs $VEV$ \hspace*{1.9cm} $\nu_{t}$ and
the mass parameter $m_{\pi_{t}}$. \vspace*{0.5cm}

\begin{figure}[htb]
\vspace*{-0.5cm}
\begin{center}
\epsfig{file=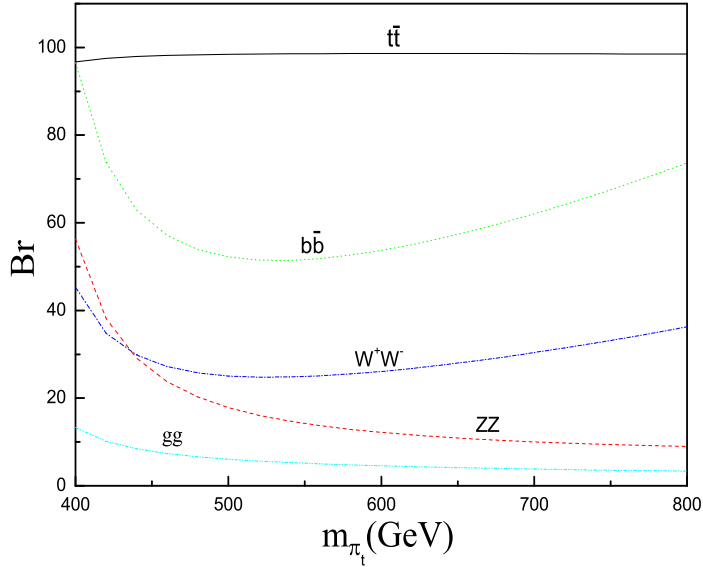,width=300pt,height=250pt} \vspace{-1cm}
\hspace{0.5cm} \caption{Branching ratios for the prominent decay
modes of the neutral top-pion $\pi^{0}_{t}$ as \hspace*{1.7cm}
functions of the mass parameter $m_{\pi_{t}}$ for
$\nu_{t}/\nu\approx1/5$.} \label{ee}
\end{center}
\end{figure}

For the charged top-pions $\pi_{t}^{\pm}$, the mainly decay mode is
$tb$. The decay width of the decay channel $\pi_{t}^{+}\rightarrow
t\overline{b}$ can be written as:
\begin{eqnarray}
\Gamma(\pi_{t}^{+}\rightarrow t\overline{b})
&=&\frac{3\alpha_{e}}{8S^{2}_{W}m^{2}_{W}m^{3}_{\pi_{t}}}
(\frac{\nu}{\nu_{t}})^{2}\lambda^{\frac{1}{2}}(m^{2}_{\pi_{t}},
m^{2}_{t},m^{2}_{b})\nonumber\\
&&[(m^{2}_{\pi_{t}}-m^{2}_{t}-m^{2}_{b})(m^{2}_{t}+m^{2}_{b})
-4m^{2}_{b}m^{2}_{t}],
\end{eqnarray}
where $\lambda(a,b,c)=(a-b-c)^{2}-4bc$. Certainly, the charged
top-pions $\pi^{\pm}_{t}$ can also decay to $WZ$ and $W\gamma$ at
loop level. However, comparing with the tree-level process
$\pi_{t}^{\pm}\rightarrow tb$, the branching ratios of the one-loop
processes $\pi_{t}^{\pm}\rightarrow WZ, W\gamma$ are very small. The
rare decay channels for the charged top-pions $\pi_{t}^{\pm}$ have
been studied in the context of topcolor-assisted technicolor(TC2)
models[10].

In the following sections, we will use above formulas to study the
production of the top-pions $\pi^{0,\pm}_{t}$ at the LHC.

\noindent{\bf III. Production of the neutral top-pion $\pi_{t}^{0}$
at the LHC}

From above discussions, we can see that the production of the
neutral top-pion $\pi_{t}^{0}$ at the LHC mainly comes from the
gluon fusion ($gg\rightarrow\pi_{t}^{0}$) and the bottom quark
fusion ($bb\rightarrow\pi_{t}^{0}$). The production cross sections
at parton level, which are proportional to the decay widths
$\Gamma(\pi_{t}^{0}\rightarrow gg)$ and
$\Gamma(\pi_{t}^{0}\rightarrow b\overline{b})$, can be approximately
written as:
\begin{equation}
\hat{\sigma}(gg\rightarrow\pi_{t}^{0})=\frac{\alpha^{2}_{s}m^{4}_{t}}
{64\pi\nu^{2}_{t}}[(1-4\frac{m^{2}_{t}}{m^{2}_{\pi_{t}}})C_{0}(m_{t}^{2})
-\frac{2}{m^{2}_{\pi_{t}}}]^{2},
\end{equation}
\begin{equation}
\hat{\sigma}(b\overline{b}\rightarrow\pi_{t}^{0}) =\frac{3\pi
m^{2}_{b}}{4\nu^{2}_{t}m_{\pi_{t}}}.
\end{equation}

The production cross section for the process
$pp\rightarrow\pi_{t}^{0}+X$ at the LHC can be expressed as:
\begin{eqnarray}
\sigma(pp\rightarrow\pi_{t}^{0}+X)&=&\int^{1}_{0}dx_{1}\int^{1}_{0}dx_{2}
f_{g/p}(x_{1},\mu_{F})f_{g/p}(x_{2},m^{2}_{\pi_{t}})\hat{\sigma}(gg\rightarrow\pi_{t}^{0})
\nonumber\\&+&\int^{1}_{0}dx_{1}\int^{1}_{0}dx_{2}
f_{b/p}(x_{1},m^{2}_{\pi_{t}})f_{b/p}(x_{2},\mu_{F})\hat{\sigma}(b\overline{b}\rightarrow\pi_{t}^{0}),
\end{eqnarray}
where $f_{i/p}$ are the parton distribution functions(PDF's) for
gluons and bottom quarks evaluated at some factorization scale
$\mu_{F}$. In our numerical calculation, we will use CTEQ6L
PDF's[11] for the quark and gluon PDF's. Following the suggestions
given by Ref.[12], we assume that the factorization scale $\mu_{F}$
for the bottom quark PDF is of order ${m}_{\pi_{t}}/4$ in this
paper.

\vspace*{0.5cm}
\begin{figure}[htb]
\hspace*{2cm} \epsfig{file=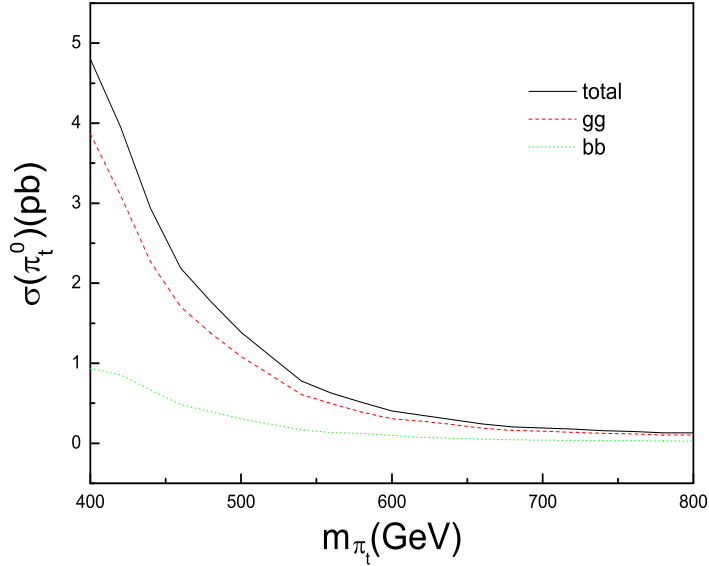,width=300pt,height=250pt}
\vspace{-1cm} \hspace{0.5cm} \caption{The production cross section
$\sigma(\pi_{t}^{0})$ at the LHC with $\sqrt{s}=14TeV$ as a
\hspace*{1.9cm} function of the mass parameter $m_{\pi_{t}}$ for
$\nu_{t}/\nu\approx1/5$.}\label{ef}
\end{figure}

The production cross section $\sigma(\pi_{t}^{0})$ of the process
$pp\rightarrow\pi_{t}^{0}+X$ at the LHC with $\sqrt{s}=14TeV$ is
plotted in Fig.2 as a function of the top-pion mass ${m}_{\pi_{t}}$
for $\nu_{t}/\nu\approx1/5$, in which the dashed line, dotted line
and solid line represent the contributions of the gluon fusion
process, bottom quark fusion process, and both these processes,
respectively. From Fig.2 we can see that the production cross
section $\sigma(\pi_{t}^{0})$ mainly comes from the parton process
$gg\rightarrow\pi_{t}^{0}$. However, the contributions of the
 process $b\overline{b}\rightarrow\pi_{t}^{0}$ to
$\sigma(\pi_{t}^{0})$ can not be neglected, which is in the range of
$935fb\sim27fb$ for $400GeV\leq m_{\pi_{t}}\leq800GeV$. This is
because the neutral top-pion $\pi_{t}^{0}$ has a enhanced coupling
with bottom quark pair,
$g_{\pi_{t}^{0}b\overline{b}}=\frac{m_{b}}{\nu_{t}}$. If we assume
the yearly integrated luminosity $\cal{L}$$_{int}=100fb^{-1}$ for
the LHC with $\sqrt{s}=14TeV$, then there will be
$1.3\times10^{4}\sim4.8\times10^{5}$ $\pi_{t}^{0}$ events to be
generated per year. Certainly, the numerical results increase as the
$VEV$ $\nu_{t}$ decreasing. For example, if we assume
$\nu_{t}=30GeV$ and $400GeV\leq m_{\pi_{t}}\leq800GeV$, there will
be $3.6\times10^{4}\sim1.3\times10^{6}$ $\pi_{t}^{0}$ events to be
generated per year at the $LHC$ with $\sqrt{s}=14TeV$ and
$\cal{L}$$_{int}=100fb^{-1}$.

\begin{figure}[htb]
\vspace*{-0.5cm}
\begin{center}
\epsfig{file=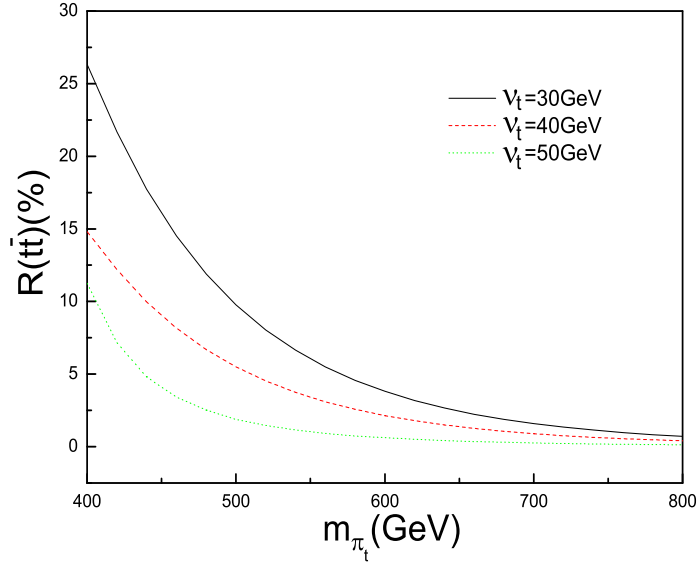,width=300pt,height=250pt} \vspace{-1cm}
\hspace{-0.5cm} \caption{The relative correction parameter
$R(t\overline{t})$ versus the mass parameter $m_{\pi_{t}}$ for
\hspace*{1.9cm} $\nu_{t}=30GeV$(solid line), $\nu_{t}=40GeV$(dashed
line), $\nu_{t}=50GeV$(dotted line).} \label{eg}
\end{center}
\end{figure}

The production cross section of the top quark pair($t\overline{t}$)
has been calculated at next-to-next-to-leading(NNL) order in the
context of the SM and can be measured to several percent accuracy at
the LHC[13,14]. Thus, it is needed to calculate the corrections of
new physics beyond the SM to the $t\overline{t}$ production cross
section. The neutral top-pion $\pi^{0}_{t}$ has large coupling with
the top quarks and should generate significant corrections to the
$t\overline{t}$ production cross section at the LHC via the enhanced
gluon and bottom quark fusion mechanisms. The relative correction
parameter
$R(t\overline{t})=\delta\sigma(t\overline{t})/\sigma^{SM}(t\overline{t})$
with
$\delta\sigma(t\overline{t})=\sigma^{HTH}(t\overline{t})-\sigma^{SM}(t\overline{t})$,
which comes from $\pi^{0}_{t}$ exchange, is plotted as a function of
the top-pion mass $m_{\pi_{t}}$ for three values of the $VEV$
$\nu_{t}$ in Fig.3. One can see from Fig.3 that the relative
correction parameter $R(t\overline{t})$ is sensitive to the mass
parameter $m_{\pi_{t}}$ and its value quickly decreases as
$m_{\pi_{t}}$ increasing. However, as long as
$m_{\pi_{t}}\leq500GeV$ and $\nu_{t}/\nu\leq1/5$, the value of
$R(t\overline{t})$ is larger than $1.9\%$, which might be detected
at the LHC experiments.

To further discuss the possible signatures of the neutral top-pion
$\pi^{0}_{t}$, we calculate the production cross section of the
process $pp\rightarrow\pi^{0}_{t}+X\rightarrow\gamma\gamma+X$ at the
LHC with $\sqrt{s}=14TeV$. Our numerical results show that the
production cross section $\sigma(\gamma\gamma)$ is very small. Even
for $m_{\pi_{t}}=400GeV$ and $\nu_{t}/\nu\leq1/8$, there will be
only about 25 $\gamma\gamma$ events to be generated per year at the
LHC with $\sqrt{s}=14TeV$ and $\cal{L}$$_{int}=100fb^{-1}$. This is
because, compared to other decay modes, the branching ratio
$Br(\pi^{0}_{t}\rightarrow\gamma\gamma)$ is very small. Thus, the
possible signals of $\pi^{0}_{t}$ predicted by the HTH model can not
be detected via the process $pp\rightarrow\pi^{0}_{t}+X
\rightarrow\gamma\gamma+X$ at the LHC experiments.

\noindent{\bf IV. Production of the charged top-pions
$\pi_{t}^{\pm}$ at the LHC}

A charged scalar does not belong to the particle spectrum of the SM.
The discovery of a charged scalar would be a clear signal for the
existence of new physics beyond the SM. At the LHC, the main
production mechanism of a light charged scalar is via the top quark
pair production processes $gg(q\overline{q})\rightarrow
t\overline{t}$ followed by the top(or anti-top) quark decay, while,
for a heavy charged scalar, the main production process at the
leading order comes from the gluon bottom quark fusion process[15].
The complete NLO QCD corrections have been calculated by Ref.[16].
In the context of the topcolor-assisted technicolor(TC2) models[17],
we have studied the production of the charged top-pions
$\pi_{t}^{\pm}$ via the parton process $gb\rightarrow
t\pi_{t}^{\pm}$ and discussed the possibility of detecting
$\pi_{t}^{\pm}$ at the LHC via the $tb(\overline{t}b)$ decay
channels[18]. The charged top-pions $\pi_{t}^{\pm}$ predicted by the
HTH model have similar features to those from the TC2 models. In the
framework of the HTH model, the production cross section
$\sigma(t\pi^{-}_{t})$ for the process $pp\rightarrow gb+X
\rightarrow t\pi^{-}_{t}+ X$ at the LHC are plotted as a function of
$m_{\pi_{t}}$ for three values of the top-Higgs $VEV$ $\nu_{t}$ in
Fig.4. For $400GeV\leq m_{\pi_{t}}\leq800GeV$ and $\nu_{t}=50GeV$,
the value of the cross section $\sigma(t\pi^{-}_{t})$ is in the
range of $1755fb\sim63fb$.

From the discussions given in section II, we can see that, for the
charged top-pion $\pi^{-}_{t}$, the main decay mode is
$\overline{t}b$. Thus, the associated production of the charged
top-pion $\pi^{-}_{t}$ with single top quark can easily transfer to
the $t\overline{t}b$ final state. Our numerical results show that,
for $m_{\pi_{t}}=600GeV$ and $\nu_{t}/\nu\approx1/5$, there will be
$2.8\times10^{4}$ $t\overline{t}b$ events to be generated per year
at the LHC with $\cal{L}$$_{int}=100fb^{-1}$. According the analysis
conclusions for the backgrounds of the process $pp\rightarrow
t\overline{t}b+ X $ given by Ref.[15,19], we expect that the charged
top-pions $\pi_{t}^{\pm}$ should be observed via the process
$pp\rightarrow t\pi_{t}^{\pm}+ X \rightarrow
t\overline{t}b(\overline{t}t\overline{b})+ X $ in the near future
LHC experiments.

\begin{figure}[htb]
\vspace*{-0.5cm}
\begin{center}
\epsfig{file=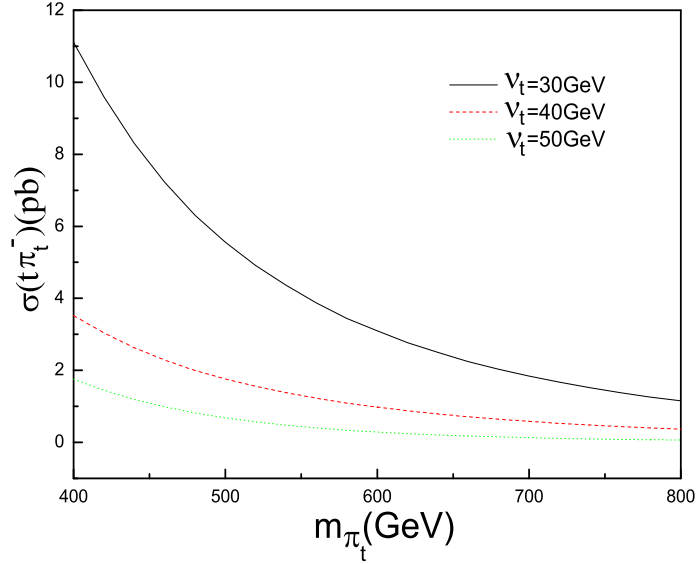,width=300pt,height=250pt} \vspace{-1cm}
\hspace{-0.5cm} \caption{The production cross section
$\sigma(t\pi^{-}_{t})$ as a function of the top-pion mass
$m_{\pi_{t}}$ for \hspace*{1.7cm} $\nu_{t}=30GeV$(solid line),
$\nu_{t}=40GeV$(dashed line), $\nu_{t}=50GeV$(dotted line).}
\label{eh}
\end{center}
\end{figure}

\begin{figure}[htb]
\vspace{-6.9cm}
\begin{center}
\epsfig{file=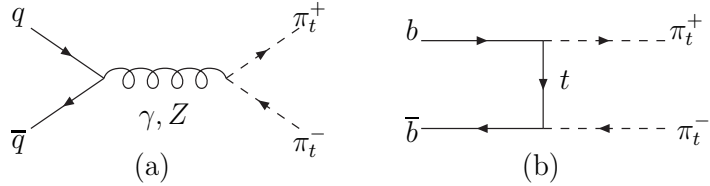,width=550pt,height=800pt} \vspace{-18.5cm}
\hspace{1cm} \vspace{-1.5cm}
 \caption{The leading order Feynman diagrams for $q\overline{q}\rightarrow
 \pi_{t}^{+}\pi_{t}^{-}$} \label{ei}
\end{center}
\end{figure}

At the LHC, the charged top-pions $\pi_{t}^{\pm}$ can also be
produced in pair production mode. The main production processes are
the usual Drell-Yan processes
$q\overline{q}\rightarrow\pi_{t}^{+}\pi_{t}^{-}(q=u,d,c,$, $s$ and $
b )$ through the s-channel Z exchange and photon exchange, and the
t-channel bottom quark scatting process
$b\overline{b}\rightarrow\pi_{t}^{+}\pi_{t}^{-}$, as shown in Fig.5.
Certainly, $\pi_{t}^{+}\pi_{t}^{-}$ production receives an
additional contribution from the $gg$ fusion process
$gg\rightarrow\pi_{t}^{+}\pi_{t}^{-}$ at one loop. However, its
contributions are very smaller than those of the tree level
processes. Thus, we will ignore the process
$gg\rightarrow\pi_{t}^{+}\pi_{t}^{-}$ in the following estimation.

Using Eq.(3) and other relevant Feynman rules, we can write the
invariant amplitude for the parton process
$q(P_{1})+\overline{q}(P_{2})\rightarrow\pi_{t}^{+}(P_{3})
+\pi_{t}^{-}(P_{4})$ as:
\begin{equation}
M=M_{s}+M_{t}
\end{equation}

For the process
$b(P_{1})+\overline{b}(P_{2})\rightarrow\pi_{t}^{+}(P_{3})
+\pi_{t}^{-}(P_{4})$, the invariant amplitude comes from Fig.5(a)
and (b):
\begin{eqnarray}
M&=&e\overline{v}(P_{2})Q\gamma_{\nu}u(P_{1})g^{\mu\nu}(P_{4}-P_{3})_{\mu}\nonumber\\
&&+\overline{v}(P_{2})\gamma_{\nu}(P_{L}g_{L}+P_{R}g_{R})u(P_{1})
\frac{g^{\mu\nu}}{\hat{s}-m_{Z}^{2}}\frac{e}{S_{W}C_{W}}(P_{4}-P_{3})_{\mu}\nonumber\\
&&+\frac{e^{2}}{2m^{2}_{W}S^{2}_{W}}(\frac{v}{v_{t}})^{2}\overline{v}(P_{2})
(m_{t}P_{R}+m_{b}P_{L})\frac{\not
q+m_{t}}{\hat{t}-m_{t}^{2}}(m_{t}P_{R}+m_{b}P_{L})u(P_{1}).
\end{eqnarray}

For $u,c,d$ and $s$ quarks, we only consider the contributions of
the s-channel process to the scattering amplitude, which can be
written as:
\begin{eqnarray}
M&=&e\overline{v}(P_{2})Q\gamma_{\nu}u(P_{1})g^{\mu\nu}(P_{4}-P_{3})_{\mu}\nonumber\\
&&+\overline{v}(P_{2})\gamma_{\nu}(P_{L}g_{L}+P_{R}g_{R})u(P_{1})
\frac{g^{\mu\nu}}{\hat{s}-m_{Z}^{2}}\frac{e}{S_{W}C_{W}}(P_{4}-P_{3})_{\mu}
\end{eqnarray}
with
\begin{eqnarray}
Q=\frac{2}{3}e,
\hspace{0.5cm}g_{L}=\frac{e}{4S_{W}C_{W}}(2-\frac{8}{3}S_{W}^{2}),\hspace{0.5cm}
g_{R}=-\frac{2eS_{W}}{3C_{W}}(q=u,c),
\end{eqnarray}
\begin{eqnarray}
Q=-\frac{1}{3}e,\hspace{0.5cm}
g_{L}=\frac{e}{4S_{W}C_{W}}(-2+\frac{4}{3}S_{W}^{2}),\hspace{0.5cm}
g_{R}=\frac{eS_{W}}{3C_{W}}(q=d,s,b).
\end{eqnarray}
Where $\hat{s}=(P_{1}+P_{2})^{2}$ and $\hat{t}=(P_{1}-P_{3})^{2}$
are the usual Mandelstam variables. We have neglected the light
quark masses in our calculation, except for the bottom quark mass in
the couplings $\pi_{t}^{\pm}tb$. Our numerical results are shown in
Fig.6, in which we plot the production cross section
$\sigma(\pi_{t}^{+}\pi_{t}^{-})$ for the process $pp\rightarrow
\pi_{t}^{+}\pi_{t}^{-}+X$ at the LHC with $\sqrt{s}=14TeV$ as a
function of the top-pion mass $m_{\pi_{t}}$ for
$\nu_{t}/\nu\approx1/5$. To comparison, we use the solid line and
dashed line to represent the contributions of the process
$q\overline{q}\rightarrow\pi^{+}_{t}\pi^{-}_{t}(q=u,c,d$ and $s)$
and the process $bb\rightarrow\pi^{+}_{t}\pi^{-}_{t}$, respectively.
From Fig.6 one can see that the production cross section of the
charged top-pion pair $\pi^{+}_{t}\pi^{-}_{t}$ mainly comes from the
usual Drell-Yan process
$q\overline{q}\rightarrow\pi^{+}_{t}\pi^{-}_{t}(q=u,c,d$ and $s )$
through the s-channel Z exchange and photon exchange. The total
production cross section $\sigma(\pi^{+}_{t}\pi^{-}_{t})$ is in the
range of $12.5fb\sim4.4fb$ for $400GeV\leq m_{\pi_{t}}\leq800GeV$
and $\nu_{t}/\nu\approx1/5$. The charged Higgs boson pair production
at the LHC has been calculated to next-to-leading order[20]. They
have shown that the total cross section for the process
$pp\rightarrow H^{+}H^{-}+X$ is smaller than $10fb$ for
$\tan\beta\leq30$ and $m_{H}\geq400GeV$. Thus, we expect that the
charged top-pions $\pi_{t}^{\pm}$ predicted by the HTH model can be
more easy detected at the LHC via this process than that for the
charged Higgs bosons $H^{\pm}$.

\begin{figure}[htb]
\vspace*{-0.5cm}
\begin{center}
\epsfig{file=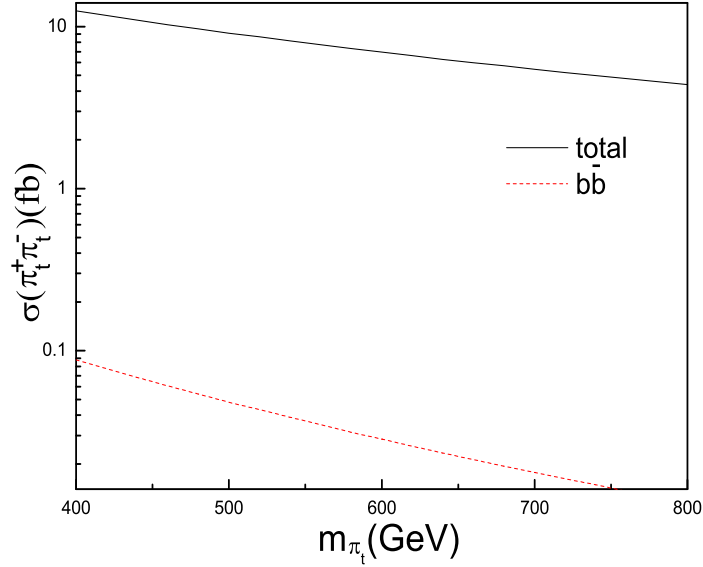,width=300pt,height=250pt} \vspace{-1cm}
\hspace{-0.5cm} \caption{The cross section
$\sigma(\pi^{+}_{t}\pi^{-}_{t})$ as a function of the top-pion mass
$m_{\pi_{t}}$ for $\nu_{t}/\nu\approx1/5$.} \label{ej}
\end{center}
\end{figure}

\noindent{\bf V. Conclusions and discussions}

Since extra dimensions can play an active role in the physics of the
TeV scales, theories with extra dimensions have recently attracted
enormous attention[21]. Based on $SU(2)\times SU(2)\times U(1)$
gauge theory in slices of two "back-to-back" $AdS'_{5}s$, topcolor
theories were reconsidered in extra dimensions[5]. For extra
dimensional descriptions of topcolor theories, there are three
possibilities for EWSB: Higgs--top-Higgs, higgsless--top-Higgs, and
higgsless--higgsless. The higgsless--top-Higgs(HTH) model which is
similar to TC2 models, predicts the existence of the top-pions
$\pi_{t}^{0,\pm}$ and the top-Higgs boson $h^{0}_{t}$. Since the
top-Higgs boson $h^{0}_{t}$ is mainly responsible for generating the
large top mass and gives small contributions to EWSB, these new
particles($\pi_{t}^{0,\pm}, h^{0}_{t}$) have large Yukawa couplings
to the third generation quarks, which might produce characteristic
signatures at various high-energy collider experiments.

In this paper, we discuss the production of the top-pions
$\pi_{t}^{0,\pm}$ predicted by the HTH model at the LHC with
$\sqrt{s}=14TeV$ and $\cal{L}$$_{int}=100fb^{-1}$ via various
suitable mechanisms (gluon-gluon fusion, bottom-bottom fusion,
gluon-bottom fusion, and the usual Drell-Yan processes). The
following conclusions are obtained.

(1)  The neutral top-pion $\pi_{t}^{0}$ can be abundantly produced
via the gluon-gluon fusion and bottom-bottom fusion processes. The
production cross section $\sigma(\pi_{t}^{0})$ is sensitive to the
top-pion mass $m_{\pi_{t}}$ and the top-Higgs $VEV$ $\nu_{t}$, which
mainly comes from the parton process $gg\rightarrow\pi_{t}^{0}$. For
$\nu_{t}\approx50GeV$ and $400GeV\leq m_{\pi_{t}}\leq800GeV$, the
value of $\sigma(\pi_{t}^{0})$ is in the range $130fb\sim4800fb$.

(2)  The main decay modes of the heavy neutral top-pion
$\pi_{t}^{0}$ are $t\overline{t}$ and $gg$, thus the neutral
top-pion $\pi_{t}^{0}$ can generate significant corrections to the
$t\overline{t}$ production at the LHC via the process
$pp\rightarrow\pi_{t}^{0}+X\rightarrow t\overline{t}+X$. For
$\nu_{t}=40GeV$ and $400GeV\leq m_{\pi_{t}}\leq600GeV$, the value of
the relative correction parameter $R(t\overline{t})$ is in the range
of $2.1\%\sim14.8\%$.

(3)  For a heavy charged scalar, the dominant production process at
the LHC is its associated production with a top quark via gluon
bottom fusion. The LHC has good potential for discovering a heavy
charged scalar through this process. Thus, we estimate the
production cross section of the process $pp\rightarrow
t\pi_{t}^{-}+X$ at the LHC with $\sqrt{s}=14TeV$ and
$\cal{L}$$_{int}=100fb^{-1}$. Our numerical results show that the
production rate is significantly large and the possible signals of
$\pi_{t}^{\pm}$ might be detected at the LHC through the process
$pp\rightarrow t\pi_{t}^{\pm}+X$ in their
$t\overline{b}(\overline{t}b)$ decay modes.

(4)  As long as the charged scalar is not too heavy, it is possible
to be produced in pair at the near future LHC experiments. The main
production processes are the usual Drell-Yan processes and the
$t$-channel bottom quark scattering process. We further estimate the
rate of the charged top-pion pair production at the LHC. Our
numerical results show that the production cross section
$\sigma(\pi_{t}^{+}\pi_{t}^{-})$ is in the range of $9.1fb\sim
18.6fb$ for $\nu_{t}=30GeV$ and $400GeV\leq m_{\pi_{t}}\leq800GeV$,
which might be larger than that for the charged Higgs bosons
$H^{\pm}(m_{H}\geq400GeV$ and $\tan\beta\leq30)$ predicted the
minimal supersymmetric standard model.

In conclusion, as long as the top-pions $\pi^{0,\pm}_{t}$ are not
too heavy($m_{\pi_{t}}<1TeV$), they can be abundantly produced at
the LHC via various suitable mechanisms. The possible signatures of
these new particles might be detected at the LHC experiments.

At last, we have to say that all of our numerical results are
obtained at leading order. The strong Yukawa couplings can give
large corrections to these numerical results. For example, for the
process $pp\rightarrow gb+X\rightarrow t\pi_{t}^{-}+X$, the leading
order cross section given in section IV can be enhanced about $80\%$
by the effects of next leading corrections. Thus, the numerical
results given in this paper can only be taken as order-of-magnitude
guides. However, we expect that our estimations can provide the
right qualitative picture.

\vspace{0.5cm} \noindent{\bf Acknowledgments}

This work was supported in part by Program for New Century Excellent
Talents in University(NCET-04-0290), the National Natural Science
Foundation of China under the Grants No.10475037.


\vspace{1.0cm}


\newpage
\begin{thebibliography}{99}
\bibitem{y1}C. T. Hill and E. H. Simmons, {\em Phys. Rept.} {\bf
             381}(2003)235; [{\em Erratum-ibid}, {\bf 390}(2004)553].
\bibitem{y2}N. Arkani-Hamed, S. Dimopoulos and G. R. Dvali, {\em Phys. Lett. B}{\bf 429}(1998)263;
            I. Antoniadis, N. Arkani-Hamed, S. Dimopoulos
            and G. R. Dvali, {\em Phys. Lett. B}{\bf 436}(1998)257;
            K. R. Dienes, E. Dudas and T.Gherghetta,
            {\em Phys. Lett. B}{\bf 436}(1998)55; {\em Nucl. Phys. B}{\bf 537}(1999)47.
\bibitem{y3}Y. Kavamura, {\em Prog. Theor. Phys.} {\bf 103}(2000)613;
            {\em Prog. Theor. Phys.} {\bf 105}(2001)691;
            A. Hebecker and J. March-Russell, {\em Nucl. Phys. B}{\bf
            613}(2001)3; {\em Nucl. Phys. B}{\bf 625}(2002)128;
            L. J. Hall, Y. Nomura, T. Okui and D. R. Smith, {\em Phys. Rev.
            D}{\bf 65}(2002)035008; L. J. Hall and Y. Nomura, {\em Phys. Rev.
            D}{\bf 65}(2002)125012.
\bibitem{y4}C. Csaki, C. Grojean, L. Pilo and J. Terning, {\em Phys. Rev.
            Lett. } {\bf 92}(2004)101802; Y. Nomura, {\em JHEP} {\bf
            0311}(2003)050; R. Barbieri, A. Pomarol and R. Rattazzi, {\em Phys. Lett. B}{\bf 591}(2004)141;
            C. Csaki, C. Grojean, J. Hubisz, Y.
            Shirman and J. Terning, {\em Phys. Rev. D}{\bf 70}(2004)015012.
\bibitem{y5}G. Cacciapaglia et al., {\em Phys. Rev. D}{\bf 72}(2005)095018.
\bibitem{y6}N. Arkani-Hamed, H. C. Cheng, B. A. Dobrescu and L. J.
            Hall, {\em Phys. Rev. D} {\bf 62}(2000)096006; M.
            Hashimoto and D. K. Hong, {\em Phys. Rev. D}{\bf 71}(2005)056004.
\bibitem{y7}G. Passarino and M. Veltman, {\em Nucl. Phys. B}{\bf 160}(1997)151;
            A. Axelrod, {\em Nucl. Phys. B}{\bf 209}(1982)349;
            M. Clements et al., {\em Phys. Rev. D}{\bf 44}(1983)570.
\bibitem{y8}The CDF Collaboration, the D0 Collaboration, the
            Tevatron Electroweak Working Group, {\em Combination of CDF
            and D0 Result on the Top-Quark mass}, {\em
            hep-ex}/0507091.
\bibitem{y9}S. Eiddman et al. [Particle data Group], {\em Phys. Lett. B}{\bf592}
            (2004)1.
\bibitem{y10}X. L. Wang, W. N. Xu, L. L. Du, {\em Commun. Theor. Phys.} {\bf
             41}(2004)737.
\bibitem{y11}J. Pumplin et al. {\em JHEP} {\bf0207}(2002)012; D. Stump et al.,
             {\em JHEP} {\bf0310}(2003)046.
\bibitem{y12}T. Plehn, {\em Phys. Rev. D}{\bf 67}(2003)014018; F.
             Maltoni, Z. Sullivan and S. Willenbrock, {\em Phys. Rev. D}{\bf
             67}(2003)093005; R. V. Harlander and W. B. Kilgore, {\em Phys. Rev. D}{\bf 68}(2003)013001;
             A. Alves and T. Plehn, {\em Phys. Rev. D}{\bf71}(2005)115014.
\bibitem{y13}ATLAS Collaboration, Technical Design Report, {\em CERN-LHCC}-99-15;
             CMS Collaboration, Technical Proposal, {\em CERN-LHCC}-94-38; G. Weiglein et al.
             [LHC/LC Study Group], {\em hep-ph}/0410364.
\bibitem{y14}M. Beneke et al., {\em "Top Quark Physics"}, {\em
             hep-ph}/0003033, D. Chakrahorty, J. Konigsherg and D.
             Rainwater, {\em Ann. Rev. Nucl. Part. Sci.} {\bf 53}(2003)301;
             W. Wagner. {\em Rept. Prog. Phys.} {\bf 68}(2005)2409.
\bibitem{y15}A. C. Bawa, C. S. Kim and A. D. Martin, {\em Z. Phys. C}{\bf
             47}(1990)75; J. F. Gunion, {\em Phys. Lett. B}{\bf
             322}(1994)125; V. D. Barger, R. J. N. Phillips and D. P.
             Roy, {\em Phys. Lett. B}{\bf 324}(1994)236.
\bibitem{y16}S. H. Zhu, {\em Phys. Rev. D}{\bf 67}(2003)075006; T.
             Plehn, {\em Phys. Rev. D}{\bf 67}(2003)014018; E. L.
             Berger, T. Han, J. Jiang, T. Plehn, {\em Phys. Rev. D}{\bf
             71}(2005)115012.
\bibitem{y17}C. T. Hill, {\em Phys. Lett. B}{\bf 345}(1995)483;
             K. D. Lane and E. Eichten, {\em Phys. Lett. B} {\bf 352}(1995)382;
             K. D. Lane, {\em Phys. Lett. B}{\bf 433}(1998)96;
             G. Cvetic, {\em Rev. Mod. Phys.} {\bf 71}(1999)513.
\bibitem{y18}Chong-Xing Yue, Zheng-Jun Zong, Li-Li Xu and Jian-Xing
             Chen, {\em Phys. Rev. D}{\bf 73}(2006)015006.
\bibitem{y19}N. Kidonakis, R. Vogt, {\em Int. J. Mod. Phys.A}{\bf 20}(2005)3171;
             D. P. Roy, {\em hep-ph}/0510070.
\bibitem{y20}J. F. Gunion et al., {\em Nucl. Phys. B}{\bf
             294}(1987)621; A. Krause, T. Plehn, M. Spira and P. M.
             Zerwas, {\em Nucl. Phys. B}{\bf519}(1998)85; A. A.
             Barrientos Bendezu and B. A. Kniehl, {\em Nucl. Phys. B}{\bf
             568}(2000)305; S. Moretti and J. Rathsman, {\em Nucl. Phys. J. C}{\bf
             33}(2004)41; H. S. Hou et al., {\em Phys. Rev. D}{\bf
             71}(2005)075014; A. Alves and T. Plehn, {\em Phys. Rev. D}{\bf
             71}(2005)115014.
\bibitem{y21}C. Csaki, {\em hep-ph}/0404096; C. Csaki, J. Hubisz and
             P. Meade, {\em hep-ph}/0510275.







\end{thebibliography}
\end{document}